\documentclass[aps,pra,amsmath,amssymb,twocolumn,superscriptaddress,showpacs,floatfix]{revtex4-1}
\usepackage{ifpdf}
 \newif\ifpdf
\ifx\pdfoutput\undefined
   \pdffalse
\else
   \pdfoutput=1
   \pdftrue
\fi
\ifpdf
   \usepackage{graphicx}
   \usepackage{epstopdf}
   \usepackage{color}
   \usepackage{verbatim}
\else
   \usepackage{graphicx}
\fi
\usepackage{bm}
\usepackage{srctex}
\usepackage{hyperref}

\begin{document}

\date{\today}

\title{Magnetic spiral order in the square-lattice spin system (CuBr)Sr$_{2}$Nb$_{3}$O$_{10}$}

\author{A.~V.~Mikheyenkov}
\affiliation{Institute for High Pressure Physics, Russian Academy of Sciences, Moscow (Troitsk) 108840, Russia}
\affiliation{Department of Theoretical Physics, Moscow Institute of Physics and Technology (State University), Moscow 141700, Russia}
\affiliation{National Research Center ``Kurchatov Institute'', Moscow 123182, Russia}

\author{V.~E.~Valiulin}
\affiliation{Institute for High Pressure Physics, Russian Academy of Sciences, Moscow (Troitsk) 108840, Russia}
\affiliation{Department of Theoretical Physics, Moscow Institute of Physics and Technology (State University), Moscow 141700, Russia}
\affiliation{National Research Center ``Kurchatov Institute'' - ITEP, Moscow 117218, Russia}

\author{A.~F.~Barabanov}
\affiliation{Institute for High Pressure Physics, Russian Academy of Sciences, Moscow (Troitsk) 108840, Russia}

\date{\today}
\begin{abstract}
\noindent
We address quantum spin helical states in the strongly frustrated Heisenberg model. Contrary to conventional Dzyaloshinskii-Moriya approach we show that such states appear without central symmetry breaking. As an example, we  demonstrate that the magnetic and thermodynamic properties of the quasi-two-dimensional square-lattice compound $\mathrm{(CuBr)Sr_{2}Nb_{3}O_{10}}$ can be interpreted within 2D $S = 1/2$ $J_1-J_2-J_3$ Heisenberg model. In this compound neutron experiment indicates helical spin order while
central symmetry does hold.
\end{abstract}

\maketitle

\section{Introduction}

\label{intro}
Helical (spiral) spin states constitute a topical and intriguing field of magnetism being the subject of intense research last years. Most of the investigations consider the helical state caused by the Dzyaloshinskii-Moriya interaction (DMI) \cite{Dzialo57_SPJ,Moriya60_PR}.

The DMI is widely used mechanism for theoretical description of neutron scattering experimental data for complex spin structures \cite{Djurek15_JMMM,Feng10_JMMM,Povzne18_PB}.
The DMI can induce helical or cycloidal magnetic structure with a determined chirality such as skyrmion with exotic thermodynamic properties \cite{Kwon14_JMMM,Okumur18_PB,Hui17_PB}. In addition, the DMI plays a key role in the ground-state phase diagram of a spin-1 Heisenberg-Ising alternating chains and appearance of the Haldane phase in such systems \cite{Liu16_JMMM}.

It is however noteworthy that DMI approach presumes broken inversion symmetry.
Basically, there exists alternative way to get helical states that does not require inversion symmetry breaking. It appears to be strongly frustrated Heisenberg model~\cite{Felcy16_PB,Schmid17_PR,Lima17_PB}. In particular, in two dimensions for the square lattice helices emerge when exchange interaction on three nearest coordination spheres are considered.

Hereinafter we address quasi-two-dimensional compound with stacked square lattice magnetic planes $\mathrm{(CuBr)Sr_{2}Nb_{3}O_{10}}$. Neutron scattering experiment~\cite{Yusuf11_PRB,Ritter13_PRB}
indicates helical spin order in this substance, while DMI-based explanation is unacceptable (inversion symmetry is preserved). This dyad brings forth the assumption to describe the magnetic order via $J_1-J_2-J_3$ Heisenberg model \cite{Yusuf11_PRB,Ritter13_PRB}.

In the present communication we verify the mentioned assumption, we show that the experimental properties of $\mathrm{(CuBr)Sr_{2}Nb_{3}O_{10}}$ \cite{Yusuf11_PRB,Ritter13_PRB,Tsujim07_JPSJ}
are well reproduced within the limits of quantum $S=1/2$ $J_1-J_2-J_3$ Heisenberg model on the square lattice.

Our consideration is strictly two-dimensional, hence at nonzero temperature long-range order is impossible due to Mermin--Wagner theorem. So we address spin-liquid state, in particular with helicoidal structure of short-range order. This approach leads to adequate description of both neutron scattering experiment and thermodynamic properties. At the very end we propose the possible way of experimental verification of the approach adequacy by the analysis of spin excitation spectrum.

The rest of the paper is organized as follows. In Section~2 we introduce the model and briefly discuss the adopted method. In Section~3 the theoretical conclusions are presented and discussed with respect to the neutron diffraction, magnetic susceptibility and specific heat experimental data for the square-lattice quasi-two-dimensional $\mathrm{(CuBr)Sr_{2}Nb_{3}O_{10}}$.
To the end, in Section~4 the results obtained are summarized.

\begin{figure}
\begin{center}
\includegraphics[width = .8\columnwidth]{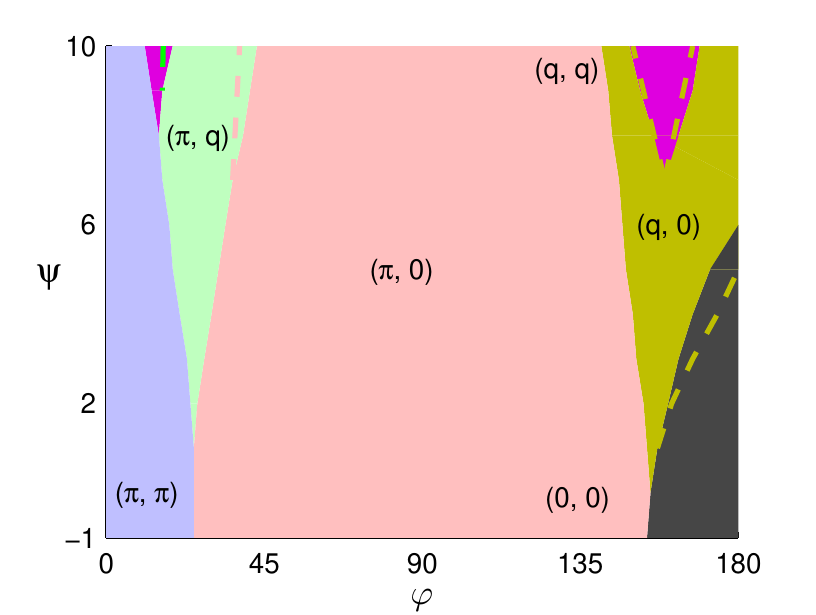}
\caption{(Color online) The regions of different short range order structure. Here $J_{1}$, $J_{2}$ and $J_{3}$ are parameterized by the angles $\varphi$ and $\psi$ (in degrees):
$J_1 = \cos \psi \cos \varphi$, $J_2 = \cos \psi \sin \varphi$,
$J_3 = \sin \psi$.
The positions of of structure factor $c_{\mathbf{q}}$ maximum in the Brillouin zone are marked. Solid borders correspond to temperature $T=0.4$, dashed lines --- $T=0.2$ \cite{Mikhee18_JETP}.
} \label{fig:PD_quant}
\end{center}
\end{figure}

\section{Model and method}
\label{sec:1}
The Hamiltonian of the model reads
\begin{equation}
\widehat{H}=J_{0}\ \widehat{h}
\label{Hamilt0}
\end{equation} \begin{equation}
\widehat{h}=J_{1}\sum_{\langle\mathbf{i},\mathbf{j}\rangle}
\widehat{\mathbf{S}}_{\mathbf{i}}\widehat{\mathbf{S}}_{\mathbf{j}}+
J_{2}\sum_{[\mathbf{i},\mathbf{j}]}\widehat{\mathbf{S}}_{\mathbf{i}}
\widehat{\mathbf{S}}_{\mathbf{j}}+
J_{3}\sum_{\{\mathbf{i},\mathbf{j}\}}\widehat{\mathbf{S}}_{\mathbf{i}}
\widehat{\mathbf{S}}_{\mathbf{j}}
\label{Hamilt}
\end{equation}
where $J_{0}$ defines the energy scale, and
$\sqrt{J_{1}^{2}+J_{2}^{2}+J_{3}^{2}} = 1$
in the dimensionless Hamiltonian~(\ref{Hamilt}),
$(\widehat{\mathbf{S}}_{\mathbf{i}})^2=3/4$,
$\langle\mathbf{i},\mathbf{j}\rangle$ denotes NN (nearest neighbor) bonds,
$[\mathbf{i},\mathbf{j}]$ denotes NNN (next-nearest neighbor) bonds and $\{\mathbf{i},\mathbf{j}\}$ denotes NNNN (next-to-next-nearest neighbor) bonds of the square lattice sites $\mathbf{i},\mathbf{j}$.

The theoretical approach adopted hereinafter is the spherically symmetric self-consistent approach (SSSA) --- spin-rotation-invariant Green's function method (RGM in alternative notation) \cite{Shimah91_JPSJ,Suzuki94_JPSJ,Junger04_PRB}.

The SSSA proved to be the appropriate for low dimension. It preserves the spin SU(2) and translation symmetries of the Hamiltonian and allows:

\textbf{i.} to satisfy the Marshall and Mermin-Wagner theorems

\textbf{ii.} to describe at $T=0$ the system states both with and without long-range order

\textbf{iii.} to find the microscopic characteristics such as the spin-excitation spectrum $\omega(\mathbf{q},T)$, the spin-gaps and the explicit form of the  dynamic susceptibility $\chi(\mathbf{q},\omega,T)$; to go beyond the mean-field approximation by introducing damping
\cite{Baraba11_TMP}.

Let us mention that SSSA has been applied to non-square lattice geometry, $S>1/2$ \cite{Rubin12_PLAa}, the systems with the anisotropic spin exchange \cite{Vladim14_EPJB} and to the doped 2D antiferromagnets \cite{Baraba01_JL}.

It is clear from the above, that SSSA is unappropriate for the description of broken symmetry states, such as box of columnar phase as well as fractionalized excitations (see e.g. \cite{Dalla14_NP} and references  therein).

In brief, the SSSA amounts to the decoupling the chain of the equations of motion for the spin Green’s function
\begin{equation}
G_{\mathbf{nm}}=\langle S_{\mathbf{n}}^{z}|S_{\mathbf{m}}^{z}\rangle
_{\omega +i\delta }=-i\int\limits_{0}^{\infty }dt\,e^{i\omega t}\langle
\lbrack S_{\mathbf{n}}^{z}(t),S_{\mathbf{m}}^{z}]\rangle
\end{equation}
at the second step
\cite{Shimah91_JPSJ,Baraba11_TMP,Hartel13_PRB}.
Due to the spherical symmetry, only the Green's functions diagonal with
respect to $\alpha =x,y,z$ are nonzero ($G^{zz}=G^{xx}=G^{yy}=G)$,
mean cite spin is zero $\langle S_{\mathbf{n}}^{\alpha }\rangle  =0$,
therefore the spin order is characterized by spin--spin correlators.
There are three branches of spin excitations degenerate with
respect to $\alpha $.

After the Fourier transformation
$S_{\mathbf{q}}^{z}=\frac{1}{\sqrt{N}}{\,}\sum\limits_{\mathbf{r}}
e^{-i\mathbf{qr}}S_{\mathbf{r}}^{z}$
Green's function $G(\mathbf{q},\omega,T)=\langle
S_{\mathbf{q}}^{z}|S_{-\mathbf{q}}^{z}\rangle _{\omega }=-\chi
(\mathbf{q},\omega,T)$
has the form
\begin{equation}
G(\mathbf{q},\omega,T)=\frac{F_{\mathbf{q}}}{(\omega ^{2}-\omega
_{\mathbf{q}}^{2})}  \label{a_GFmf1},
\end{equation}
see \cite{Baraba11_TMP,Mikhey11_JL}
for calculational details and the cumbersome expressions for
$F_{\mathbf{q}}$ and the spin excitations spectrum $\omega_{\mathbf{q}}$
(both  --- $T$-depending).
For $J_{1}-J_{2}-J_{3}$ model the Green's function $G(\mathbf{q},\omega,T)
$ involves the correlators $c_{\mathbf{r}}$ for the first eight coordination
spheres. The correlators are to be evaluated self-consistently in terms of
$G(\mathbf{q},\omega,T)$. In addition, $G(\mathbf{q},\omega,T)$ must satisfy
the spin constraint $c_{\mathbf{r=0}}=\left\langle
S_{\mathbf{i}}^{z}S_{\mathbf{i}}^{z}\right\rangle =1/4$\ (sum
rule). These conditions are
\begin{equation}
c_{\mathbf{r}}=\frac{1}{N}\sum_{\mathbf{q}}c_{\mathbf{q}}e^{i\mathbf{qr}}; \label{a_Cr_1}
\end{equation} \begin{equation}
c_{\mathbf{r=0}} =
\frac{1}{N}\sum_{\mathbf{q}}c_{\mathbf{q}} = 1/4; \label{a_constr_1}
\end{equation}
where the structure factor $c_{\mathbf{q}}$
\begin{multline}
c_{\mathbf{q}}=\left\langle
S_{\mathbf{q}}^{z}S_{\mathbf{-q}}^{z}\right\rangle =\\
= -\frac{1}{\pi}
\int_{0}^{\infty }d\omega \,\left(
2m(\omega,T)+1\right) \mathrm{Im}\,G(\mathbf{q},\omega,T) ;
\label{a_Cq_1}
\end{multline}
and $m(\omega,T)$ stands for Bose function.
Note that eq. (\ref{a_constr_1}) fixes a vertex correction, introduced at the decoupling of many-particle Green’s functions, see, e.g. \cite{Mikhey13_JL,Mikhey16_JMMM} for details.
Note also, eq. (\ref{a_constr_1}) means that spin constraint $S_{\mathbf{i}}^2 = 3/4$ is hold exactly
on each site, while in alternative approaches --- Dyson-Maleev \cite{Yang97_PLA}, modified spin waves \cite{Hauke11_NJP} and Schwinger bosons \cite{Feldne11_PRB} it is fulfilled only in average.

\begin{figure}
\begin{center}
\includegraphics[width = .8\columnwidth]{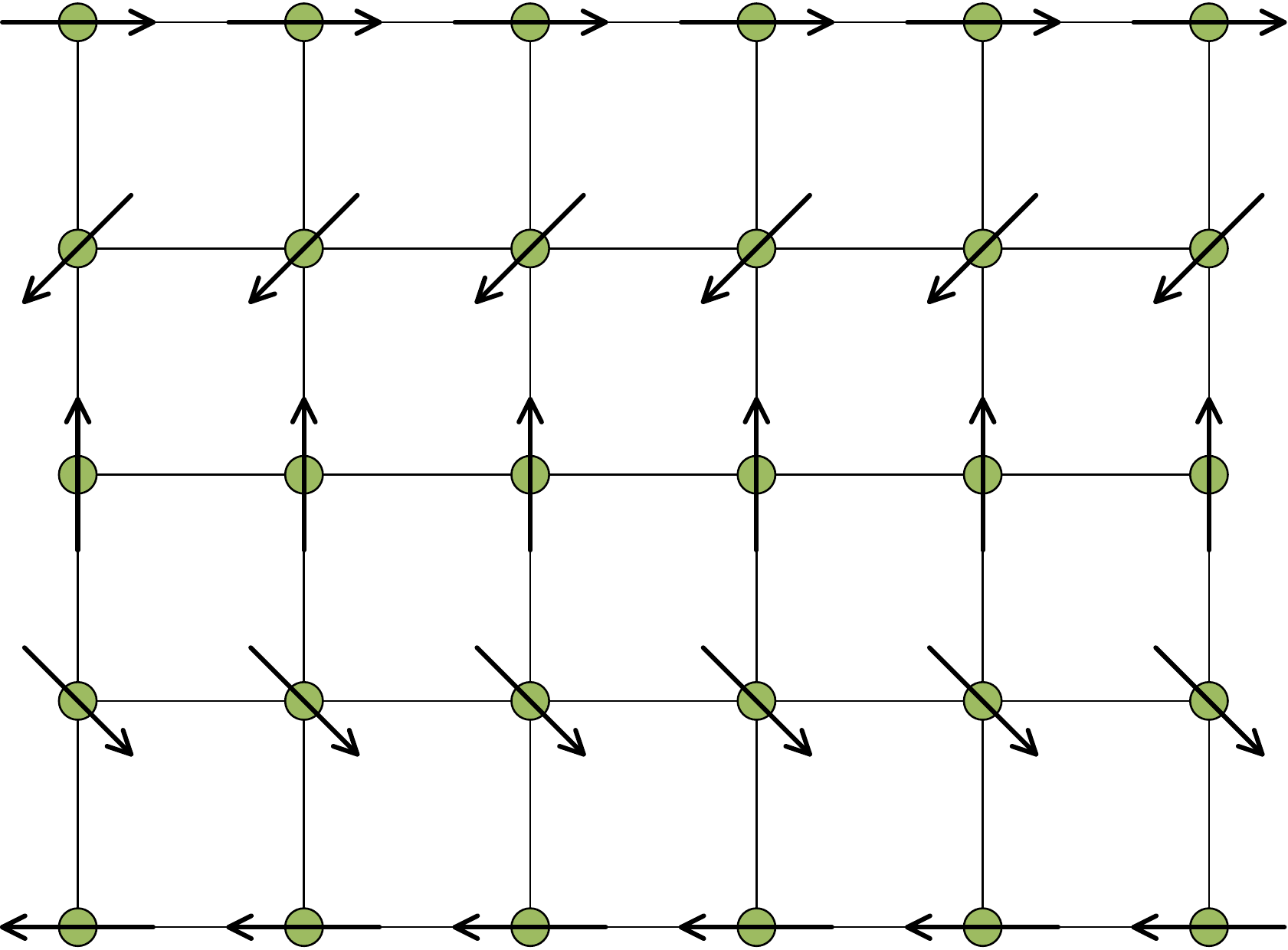}
\caption{(Color online) The spin helical structure for one plane of layered compound $\mathrm{(CuBr)Sr_{2}Nb_{3}O_{10}}$~\cite{Yusuf11_PRB}.
The helix corresponds to the structure factor $c_{\mathbf{q}}$ maximum position at the point $(3\pi /4, 0)$ in the Brillouin zone.
} \label{fig:Spins_exper}
\end{center}
\end{figure}

The expression (\ref{a_Cr_1}) generates eight equations for $\mathbf{r}$
belonging to first eight coordination spheres. This system of self-consistent
equations supplemented by (\ref{a_constr_1}) is then solved numerically.

In the general case, both short-range order (SRO) and long-range order (LRO)
states can be realized in the described approach. Because the dimension is equal to two, only SRO are possible at $T\neq 0$, and both possibilities can take place as $T\rightarrow 0$ (LRO is characterized by nonzero spin--spin correlators at infinity
$\langle S_{\mathbf{r}}^{\alpha }S_{\mathbf{0}}^{\alpha }\rangle _{r\rightarrow \infty}$). In what follows we consider only the case of nonzero temperatures.

\section{Results and discussion}
\label{sec:2}
The aim of the present work is to describe the experimental data for the layered square-lattice compound $\mathrm{(CuBr)Sr_{2}Nb_{3}O_{10}}$
\cite{Yusuf11_PRB,Ritter13_PRB,Tsujim07_JPSJ}.
This appears to be particular case of the general picture of $J_1-J_2-J_3$
model thermodynamic properties. We first describe it briefly. We set aside layer-layer interaction and consider the problem in the purely 2D case. The phase diagram of the model obtained in SSSA for two different temperatures is presented in Fig.~\ref{fig:PD_quant}. Hereafter $J_{1}$, $J_{2}$ and $J_{3}$ are parameterized by the angles $\varphi$ and $\psi$:
$J_1 = \cos \psi \cos \varphi$, $J_2 = \cos \psi \sin \varphi$,
$J_3 = \sin \psi$.

Spin order is dictated by the position of the structure factor $c_{\mathbf{q}}$
maximum in the Brillouin zone. Note that in 2D at $T \neq 0$ LRO is absent and the correlation length is defined by $c_{\mathbf{q}}$ height. It is seen from Fig.~\ref{fig:PD_quant}, that six phases are realized --- FM (maximum $c_{\mathbf{q}}$ at $(0,0)$), AFM~($(\pi,\pi)$), stipe~($(\pi,0)$) and three types of helices~($(q,0),(q,q),(\pi,q)$), we have omitted the equivalent points.

The spin order of $\mathrm{(CuBr)Sr_{2}Nb_{3}O_{10}}$ proposed in \cite{Yusuf11_PRB},
see Fig.~\ref{fig:Spins_exper}, corresponds to $c_{\mathbf{q}}$ maximum at the point $(3\pi /4, 0)$.  We remind that in our 2D approach mean cite spin is zero, and the position of $c_{\mathbf{q}}$ maximum defines short-range order expressed through spin--spin correlators.

It is also worth noting that the position of $c_{\mathbf{q}}$ maximum does not allow to define the unique values of exchange parameters $J_1$, $J_2$, $J_3$ (apart from normalization). Any set of exchange parameters corresponds to a line on $J_2/J_1$, $J_3/J_1$ plane. We have used the exchange parameters values corresponding to point $(3\pi/4,0)$ and minimal $J_3$ (it is physically obvious that $J_3$ is weak): $J_1=-0.81$, $J_2=0.56$, $J_3=0.17$ (these values are related to parametrical angles
$\varphi = 146^{\circ}$ and $\psi = 10^{\circ}$ see Fig~\ref{fig:PD_quant}).

\begin{figure}
\begin{center}
\includegraphics[width = .8\columnwidth]{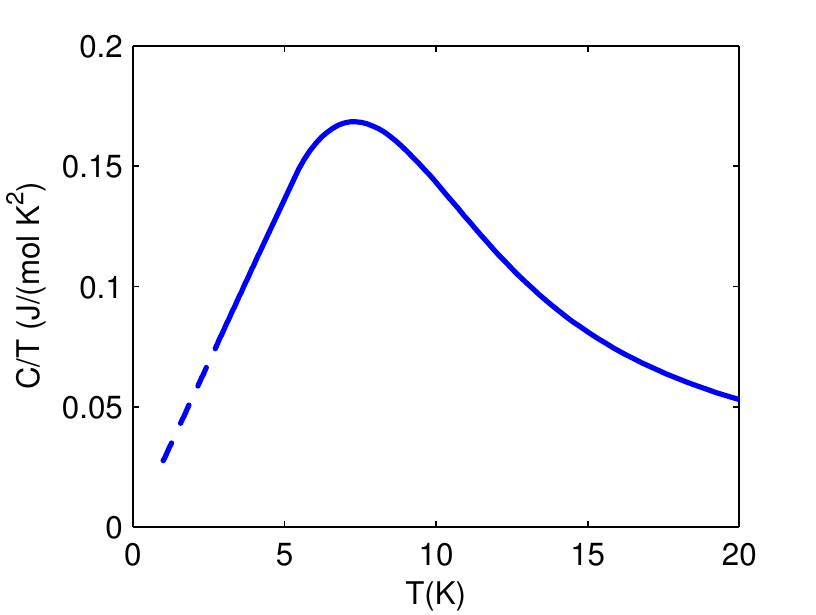}
\caption{(Color online) The calculated heat capacity for exchange parameters $J_1=-0.81$, $J_2=0.56$, $J_3=0.17$ relevant to $\mathrm{(CuBr)Sr_{2}Nb_{3}O_{10}}$. The energy scale $J_0=38K$ was defined by fitting the calculated maximum position to the experimental one.
}
\label{fig:Capacity_theor}
\end{center}
\end{figure}

\begin{figure}
\begin{center}
\includegraphics[width = .8\columnwidth]{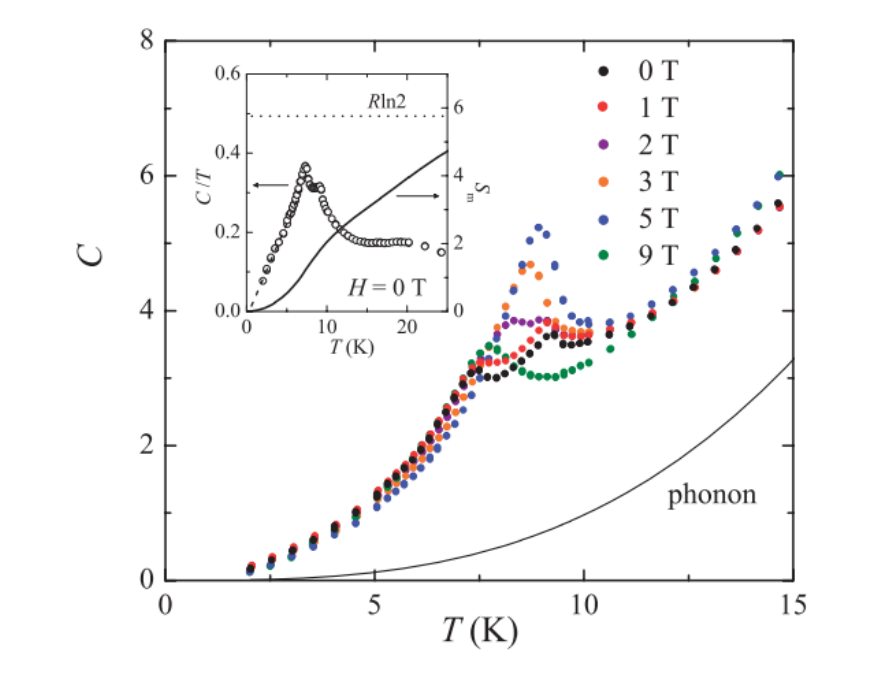}
\caption{(Color online) The experimental data for $\mathrm{(CuBr)Sr_{2}Nb_{3}O_{10}}$
heat capacity~\cite{Tsujim07_JPSJ} (with kind permission of Y. Tsujimoto).
Compare the insert (purely magnetic contribution) with the calculated curve in Fig.~\ref{fig:Capacity_theor}.
}
\label{fig:Capacity_exper}
\end{center}
\end{figure}

Fig.~\ref{fig:Capacity_theor} represents the calculated heat capacity $C/T$ vs $T$ for the discussed exchange parameters. It is to be compared with the experimental curve presented in Fig.~\ref{fig:Capacity_exper} (from \cite{Tsujim07_JPSJ}). The $T$ axes is Fig.~\ref{fig:Capacity_theor} is in Kelvins, the corresponding energy scale $J_0=38K$ was defined by fitting the calculated maximum position to the experimental one ($7.5 K$). The curves of both figures are in qualitative agreement, though there still remains quantitative factor of about two.
Note, that we use the simplest variant of SSSA --- one vertex approximation. The complificaltion of the approach (see e.g. \cite{Baraba11_TMP}) can match the experiment.

In fact the peak on Fig.~\ref{fig:Capacity_exper} has rather complex structure. The present approach does not catch fine structure of the peak. It may be the results of the interplay of 2D correlations, some spin anisotropy, the interlayer coupling, which are not taken into account in the considered model.

\begin{figure}
\begin{center}
\includegraphics[width = .8\columnwidth]{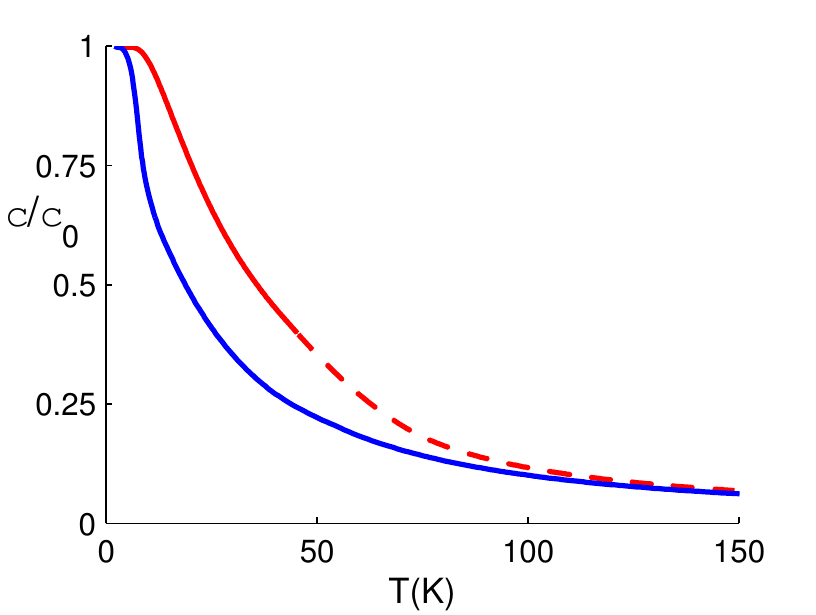}
\caption{(Color online)
Red curve -- calculated magnetic susceptibility $\chi$ normalized by $\chi_0 = \chi(T \to 0)$ for $J_1=-0.81$, $J_2=0.56$, $J_3=0.17$ relevant to $\mathrm{(CuBr)Sr_{2}Nb_{3}O_{10}}$ (the energy scale $J_0=38K$). Dashed extrapolation is the fit to Curie-Weiss law.
Blue curve --- experimental susceptibility $\chi$ (\cite{Tsujim07_JPSJ}) normalized by $\chi_0 = \chi(T \to 0)$ (data kindly granted by Y. Tsujimoto).
} \label{fig:Susc_theor}
\end{center}
\end{figure}

The calculated magnetic susceptibility $\chi(T)$ is also in a good qualitative agreement
with the experimental curve. This can be seen from Fig.~\ref{fig:Susc_theor}, where both calculated and experimental curves are presented. The calculated in the frames of the present approach
static uniform susceptibility $\chi$, as well as its fit (see \cite{Tsujim07_JPSJ})
to the Curie-Weiss law plus a T-independent term qualitatively coincide
with the measured curves for $\mathrm{(CuBr)Sr_{2}Nb_{3}O_{10}}$.
We remind that the simplest variant of SSSA is used thus its complificaltion can match the experiment.

\begin{figure}
\begin{center}
\includegraphics[width = .85\columnwidth]{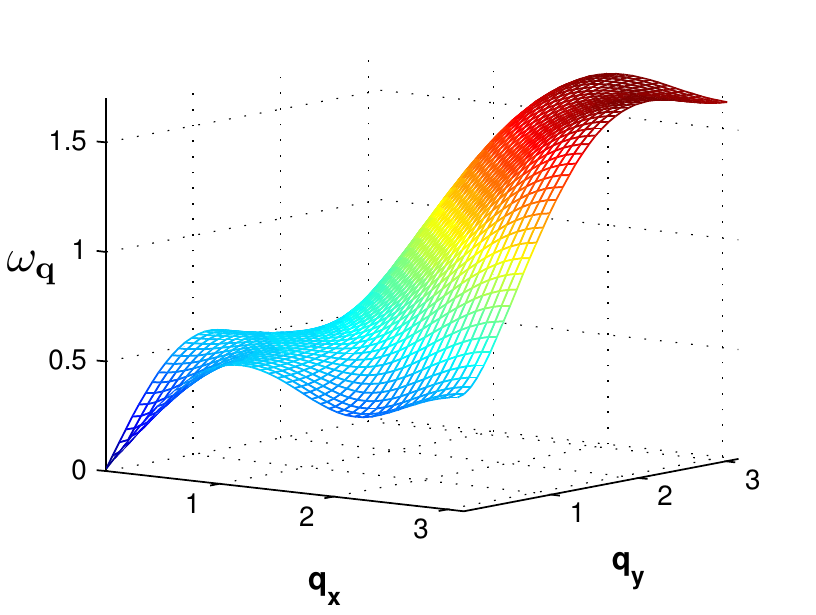}
\caption{(Color online) Spin excitations spectrum for frustrated Heisenberg model with the adopted exchange parameters $J_1=-0.81$, $J_2=0.56$, $J_3=0.17$ relevant to $\mathrm{(CuBr)Sr_{2}Nb_{3}O_{10}}$ in the frames of SSSA. Quarter of the Brillouin zone is depicted. Note large nearly dispersionless region, local minimum on the Brillouin zone side, and ``beak'' near the AFM point ($\pi,\pi$).
} \label{fig:Spectrum}
\end{center}
\end{figure}

Thus, the self consistent spherically symmetric approach for 2D frustrated $S = 1/2$ $J_1-J_2-J_3$ Heisenberg model gives a satisfactory description both neutron scattering experiment and thermodynamic properties of the quasi-two-dimensional square-lattice compound $\mathrm{(CuBr)Sr_{2}Nb_{3}O_{10}}$.
The comparison of the other theoretical results with the experiment could serve as the verification procedure. Fig.~\ref{fig:Spectrum} represents the spin excitation spectrum for the used exchange parameters. Its particular features --- large nearly dispersionless region, local minimum on the Brillouin zone side, and ``beak'' near the antiferromagnetic point ($\pi,\pi$) --- being tested experimentally could prove or discard the supposed approach.

\section{Conclusion}

To summarize we have shown that the magnetic and thermodynamic properties
of the quasi-two-dimensional square-lattice compound $\mathrm{(CuBr)Sr_{2}Nb_{3}O_{10}}$,
where neutron experiment indicates helical spin order while common explanation
by Dzyaloshinskii-Moriya interaction is unacceptable, can be interpreted
within the strongly frustrated Heisenberg model.

\section{Acknowledgements}

The authors are grateful to N.M. Chtchelkatchev for useful discussions. This work is supported by Russian Foundation for Basic Research, grant 19-02-00509. Numerical simulations were supported by Russian Science Foundation (grant RNF 18-12-00438). Part of calculations was performed using the resources of the Federal Collective Usage Center Complex for Simulation and Data Processing for Mega-science Facilities at NRC ``Kurchatov Institute'', http://ckp.nrcki.ru/, and  the cluster of Joint Supercomputing Center, Russian Academy of Sciences.

\end{document}